\documentstyle[aps,prl,multicol,epsf,epsfig]{revtex}
\begin{document}
\newcommand{\beq}{\begin{equation}}
\newcommand{\eeq}{\end{equation}}
\newcommand{\beqa}{\begin{eqnarray}}
\newcommand{\eeqa}{\end{eqnarray}}
\def\nn{\nonumber}
\def\l({\left(}
\def\r){\right)}

\title{Large scale extragalactic jets powered by very-high energy gamma rays}

\author{A.~Neronov$^a$, D.~Semikoz$^{bc}$, F.~Aharonian$^d$, O.~Kalashev$^b$}

\address{$^a$ Theoretische Physik, Universit\"at M\"unchen, Theresienstr. 37, 
80333 Munich, Germany,\\
$^b$ Institute for Nuclear Research, 60th October Anniversary prosp. 7a,
Moscow, 117312, Russia\\
$^c$ Max-Planck-Institut f\"ur Physik,
F\"ohringer Ring 6, 80805 Munich, Germany\\
$^d$ Max-Planck-Institut f\"ur Kernphysik,Saupfercheckweg 1, D-69117 Heidelberg, Heidelberg, Germany} 

\maketitle
\draft
\begin{abstract}
The radiative cooling  of electrons 
responsible for  the nonthermal synchrotron emission of 
large scale jets of radiogalaxies and quasars requires 
quasi continuous (in time and space) production  of 
relativistic electrons  throughout the jets 
over the scales exceeding 100 kpc.  While in  the standard 
paradigm  of large scale jets this implies {\em in situ} 
acceleration of electrons,  in this letter 
we propose a principally different ``non-acceleration'' origin of
these electrons, assuming  that they are implemented 
all over the length of the jet through effective development 
of electromagnetic cascades initiated by 
extremely high energy $\gamma$-rays injected into the jet 
from the central object.  This  scenario
provides a natural and very economic way to power the 
jets up to distances of 100 kpc and beyond.    
\end{abstract}


\begin{multicols}{2}

Although there is little doubt in nonthermal origin of 
radiation of large scale extragalactic jets, it remains 
a theoretical challenge to explain how the relativistic electrons 
responsible for the radio-to-X-ray synchrotron 
emission could  be distributed  quite uniformly  
all over the huge length of the jet $L_{jet}\sim 100$ kpc. 
Although the recent exciting discoveries by the Chandra
observatory added  much to our knowledge of structures of 
these fascinating components of powerful radiogalaxies
and quasars, they didn't solve the old problems and, if fact,
brought new puzzles. In particular, there remain yet 
substantial problems concerning the 
identification of  the X-ray emission  mechanism. 
The most natural mechanism, the synchrotron radiation of 
multi-TeV electrons, applied in a standard manner  to large scale 
jets, faces certain difficulties to interpret 
the observed X-ray features of a significant fraction of 
the Chandra jets (see e.g. Ref \cite{kraw_harris}).  
Therefore the recent 
``trend'' which invokes  relativistic bulk motion with Lorentz 
factor as large as $\Gamma \sim 10$ (on $\geq$ 100 kpc scales !),  
and thus increases significantly the efficiency of the inverse 
Compton scattering  on 2.7~K cosmic background radiation (CMBR) 
as a source of 
X-rays  \cite{tavecchio,celotti}, was readily accepted by the 
extragalactic jet community \cite{kraw_harris}.  The very idea of 
relativistic bulk motion could be very productive also for 
other possible jet models 
\cite{aharonian2002}. In particular,  it would allow explanation of 
the broad-band data of the jet of 3C 273 
by a single population of electrons \cite{derm_at}. 
Even so, this effect  does not solve the problem of 
continuous (in space and time)  acceleration of 
multi-TeV  electrons.  The association of the 
observed X-rays to very high-energy (VHE) protons could be 
a possible solution \cite{aharonian2002}. 

In this letter we suggest a ``non-acceleration''  version 
of the electron synchrotron model, namely assuming that  
electrons {\em are not accelerated} 
in the jet, but are result of pair production of 
extremely high energy $\gamma$-rays interacting with the CMBR . 
The  pair production in the field of CMBR and 
diffuse infrared background photons  
as a source of ultra-relativistic electrons created far from the
central object (AGN)  has been suggested \cite{halo} in the context of 
formation of giant extragalactic pair  halos. The same process 
can provide ultrarelativistic electrons for nonthermal
synchrotron X- and $\gamma$-rays in 
clusters of galaxies and beyond \cite{aharonian2002}. Recently, 
highly collimated neutral beams (neutrons and $\gamma$-rays)
have been  noticed  as possible supplier of 
synchrotron-emitting electrons in the extended jets of FRII 
radio galaxies \cite{derm_at,at_derm}.      

The mean free path of $\gamma$-rays in the field of CMBR 
has a minimum $L_\gamma\approx 8$ kpc at $E_\gamma 
\simeq 10^{15} \ \rm  eV$.  At both 
lower and higher energies $L_\gamma$ increases --  
sharply (exponentially) at $E_\gamma \ll 10^{15} \ \rm eV$
(due to the threshold effect $\gamma$-rays interact with 
the Wien tail of CMBR), and 
slowly, $L_{\gamma}
\approx\  14.6\ E_{\gamma,  16}T_{2.7}^{-2}[1+0.7\ {\rm ln}\ (E_{\gamma,16}T_{2.7})]^{-1} $ kpc 
at $E_\gamma \gg 10^{15}$ eV 
due to the decrease of the cross-section  
with the parameter  $E_\gamma   E_{\rm CMBR}\gg m^2_{\rm e}c^4$.
Hereafter, $E_{\gamma,16}=E_\gamma/10^{16}$ eV, $T_{2.7}=T/2.7 $ K. 
Thus, a $\gamma$-ray beam with a broad spectrum
extending to $10^{18} \ \rm eV$ can supply 
the jet with ``desirable'' VHE electrons along the
jet, and thus power the  
jet up to distances $\sim 100$ kpc or even more.

\begin{figure}[t]
\begin{center}
\epsfig{file=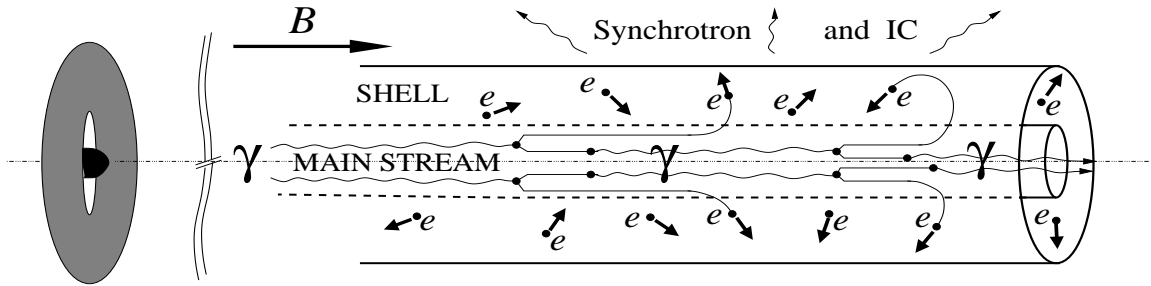,angle=-0,width =0.85\linewidth, height=3.7cm}
\end{center}
\caption{Extremely high energy gamma-rays, $E \gg 10^{14} \ \rm eV$, 
 from central engine form the  ``main stream'' of the jet and 
provide VHE electrons throughout the entire jet. The low energy electrons   
below the critical energy $E_{\rm crit}$ 
escape the ``main stream'' and form a  ``shell'' of the jet. Distant observer 
see the synchrotron and inverse Compton radiation from electrons in the "shell".}
\label{fig:cartoon}
\end{figure}

In what follows we analyze a simple model 
in which a VHE photons  are injected into a cylindrical jet and, interacting 
with CMBR as well as with the low-frequency radiation of the jet
itself, initiate an electromagnetic cascade, the relevant processes 
being  pair-production (PP), inverse 
Compton scattering (ICS) and synchrotron 
radiation  (see Fig. \ref{fig:cartoon}). 
For numerical calculations we use the code described in Ref.\cite{code} which solves kinetic 
equations   for VHE particles cascading on soft photon backgrounds.

The first generation of pair-produced electrons upscatter the soft 
background photons.  Propagation
length of electron with respect to ICS in Klein-Nishina regime  is  
$L_{e, {\rm KN}}
\approx 1.5\  T_{2.7}^{-2}E_{e,16}$ kpc. 
The electrons suffer also synchrotron losses.    

High resolution radio observations show that the magnetic 
field in large scale jets may dominate by the regular 
component 
parallel to the jet axis, although there could be anomalous regions 
near the knots with oblique field components (see e.g. Ref.\cite{regfiled}).
Here we adopt a simplified picture assuming that 
the field consists of two - random and regular 
\vskip-0.8cm
\end{multicols}
\begin{multicols}{2}
%
\vskip-0.5cm
\noindent  
(aligned with the jet axis) - components, and that  
the regular component significantly exceeds 
the random component. We normalize these fields as 
$B_0=B_{0,-5}\times  10^{-5}$ G and 
$B_r\sim 0.01 B_0 = B_{{\rm r},-7}\times 10^{-7}$ G, respectively.
Generally, if the first generation electrons appear 
at small angle to the regular magnetic field
(i.e. the $\gamma$-ray beam from the central engine is directed along the jet),
they will spiral with a small pitch angle $\theta$  
in the   magnetic field $B_0$ and at the same time get 
gradually deflected  by the random magnetic field 
$B_{\rm r}$. As long as the deflection of electrons does not 
exceed a few degree, the synchrotron losses of electrons 
are dominated by the random magnetic field. At larger angles 
the synchrotron radiation in the regular field becomes more important.
Moreover, for $\theta \geq 30^{\circ}$ the latter 
dominates also over  the Compton losses. 
At the same time, because of the Klein-Nishina effect, 
the electrons of extremely high energies, $E \geq 10^{17} \rm \ eV$ 
are basically cooled through the synchrotron radiation, 
even when they move at small pitch angles. 

Deflection of electrons in the jet  can be approximately
described as diffusion in pitch angle \cite{jokipii,gould}. The 
diffusion length  $L_{\rm diff}$ can be estimated as \cite{gould}
$
L_{\rm diff}\approx 5\ B_{{\rm r},-7}^{-2}E_{e,14}^2l_{\rm pc}^{-1}
(\theta_0 /3^{\circ})^{-2}\mbox{ kpc}~,
$
where $l_{\rm pc}$ is the correlation length of the random  
B-field  in parsecs and $\theta_0$ is the initial opening angle of the 
primary photon beam. 
When electrons cool down to  
$\sim 10^{14}$ eV, their propagation length is determined by ICS  
$
L_{e}=0.3 (T_{2.7})^{-4}(E_{e,14})^{-1}\mbox{ kpc}~.
$
When  $L_{\rm diff}$ becomes of order of $L_e$ the 
cascade electrons are  essentially  deflected by the random B-field. 
Comparing the diffusion length to the propagation elngth
we find that the electron trajectories below the energy 
$
E_{\rm crit} = 4\times 10^{13}B_{{\rm r},-7}^{2/3}
l_{\rm pc}^{1/3} T_{2.7}^{-4/3}(\theta_0 /3^{\circ})^{2/3}\mbox{ eV}
$ 
are randomized. Such electrons are cooled through  
synchrotron and ICS losses,  and form a bright "shell" 
around 
the "main stream" of  the cascade as it is 
demonstrated in Fig.~\ref{fig:cartoon}.

The energy losses  of "shell" electrons are 
dominated
\vskip-0.8cm
\begin{figure}[t]
\begin{center}
\epsfig{file=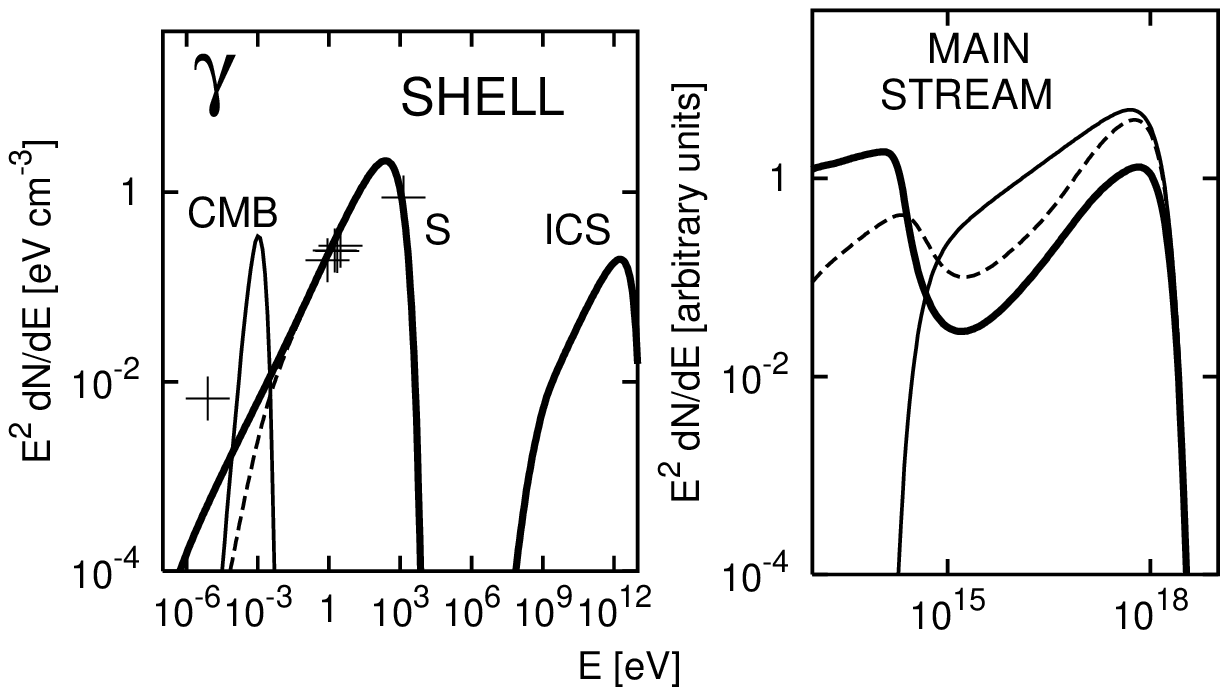,angle=-90,width =0.8\linewidth}
\epsfig{file=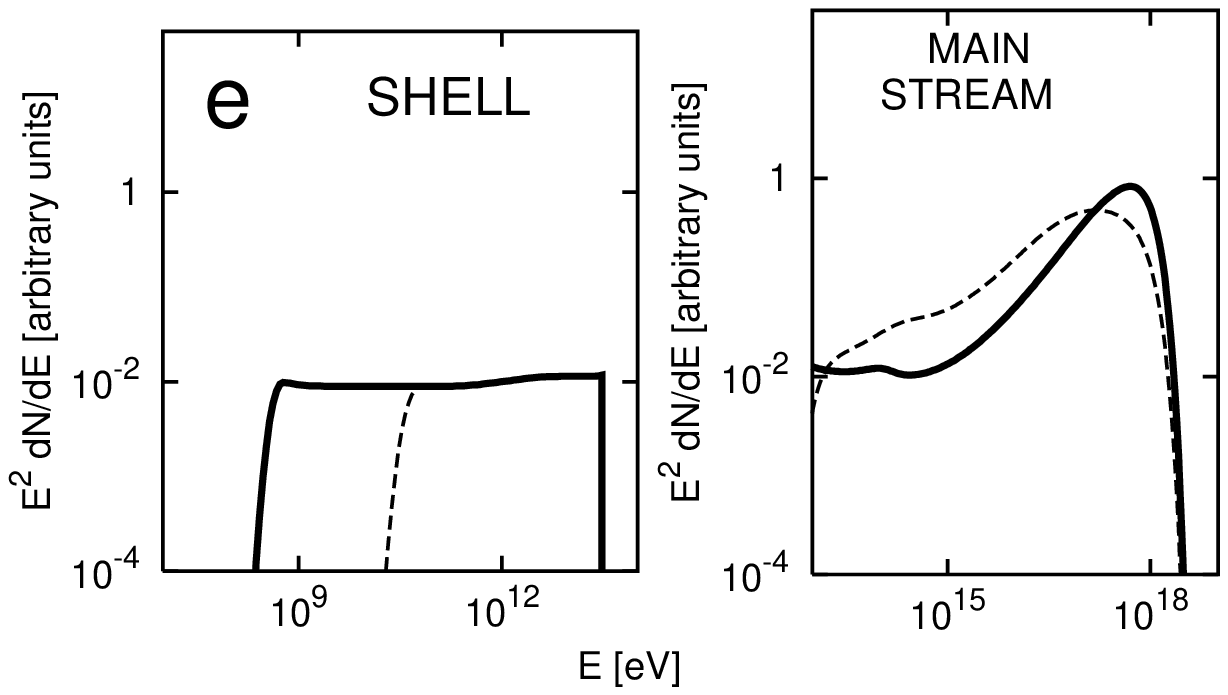,angle=-90,width =0.8\linewidth}
\caption{Evolution of energy spectra of $\gamma$-rays (upper panel) 
and electrons (lower panel) in the ``main stream'' (right) and 
in the "shell" (left) of the jet. 
Upper panel (right): thin line -- initial photons (at the base of the jet),
dashed lines -- after 50 kpc, thick lines -- 500 kpc. 
Upper panel (left):  
the  synchrotron/ICS spectra of the "shell" (dashed line -- after 
$10^6$ yr, thick solid line -- after  $10^8$ yr). For comparison 
the experimental points of the knot A of the quasar  3C 273
{\protect\cite{marshall}}  are also shown, assuming the knot size 1 kpc. 
The CMBR background  is shown by the thin solid line.
Lower panel: evolution of electron spectra  in the "main stream" and 
in the "shell" (notations are the same as in the upper panel).}
\label{fig:spectrum}
\end{center}
\end{figure}
\vskip-0.5cm
\noindent  
by the regular magnetic field which 
establish a standard $dn_{e}/dE_e \propto  E_e^{-2}$ type spectrum of
electrons. Correspondingly, the spectrum of synchrotron radiation 
behaves as $\nu F_\nu \propto  \nu^{0.5}$ with a maximum at 
$
h \nu_{\rm max} =1.6~B_{0,-5}(E_{\rm crit}/10^{14} \ \rm eV)^2 \mbox{ keV}~.
$ 
Note that at low energies a possible energy dependent-escape
might make the electron spectrum, and subsequently the synchrotron 
radio spectrum somewhat steeper. 
\vskip-0.8cm
\begin{figure}[t]
\epsfig{file=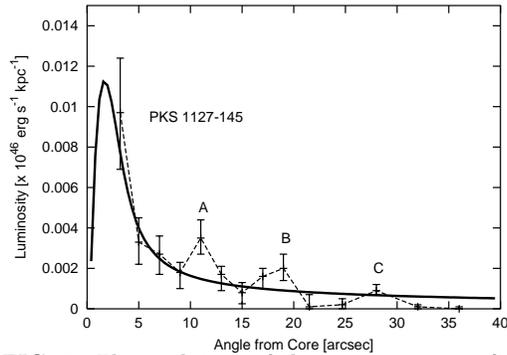,angle=-90,width =0.8\linewidth}
\caption{The evolution of the injection rate of the cascade 
electrons with energy 
$E \leq E_{\rm crit}=3 \times 10^{13} ~{\rm eV}$ to the "shell".
For comparison, the angular distribution of the X-ray 
luminosity of the jet in the quasar
PKS~1127-145  \protect\cite{pks} is shown. 
Primary spectrum of $\gamma$-rays was taken as power-law 
$E^{-1.5}$ with a cutoff at $10^{18}$ eV. Luminosity of the primary 
$\gamma$-ray beam is $10^{46}$ erg/s. Total luminosity of electrons in 
the shell is 10 \% 
of the primary beam luminosity.}
\label{fig:PKS}
\end{figure}
\vskip-0.3cm

Because the spectra of Chandra jets typically extend 
beyond 1 keV, for a reasonable value of 
$E_{\rm crit} \sim 10^{14} \rm \ eV$, which is a free parameter 
in our model, the regular B-field should exceed 
$10^{-5} \rm \ G$. This explains the need in the regular 
field in our model. We need sufficiently strong magnetic filed 
in order to explain the X-ray data. The random field cannot 
play this role because  such a strong random field would
destroy the cascade from very beginning. On the other hand we need a
cascade in order to have a more or less homogeneous distribution 
of electrons, both in density and energy spectrum,  across the jet. 
Assuming  that the maximum  energy of synchrotron radiation 
from the jet lies in the X-ray energy domain, which minimizes the energy 
requirement to the source, we find the following 
relation between the random and regular magnetic fields:~$B_{{\rm r},-7}^{2}B_{0,-5}^{3/2}l_{\rm pc}\approx 7 T_{2.7}^{4}(\theta_0/3^o
)^{-2} \left(\left. h \nu_{\rm max}\right/ {\rm 1 keV}\right)^{3/2}
$. An example of numerical calculation of evolution of photon and electron 
spectra in the "main stream" and in the "shell" are shown in  Fig. 
\ref{fig:spectrum}.

\begin{figure}[t]
\epsfig{file=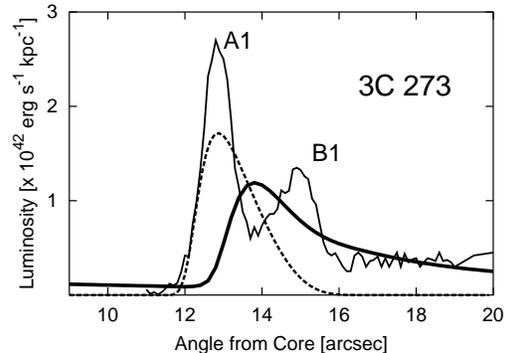,angle=-90,width =0.8\linewidth}
\caption{Evolution of the jet's luminosity. The thin solid line represents 
X-ray data for 3C 273 \protect\cite{marshall}, the dashed line 
is intensity profile of the synchrotron background 
used in numerical calculation. Thick solid line is the 
injection rate of electrons with energy 
$E \leq E{\rm crit}=10^{14}$ eV from the "main stream" of 
the electromagnetic  cascade, i.e. the rate of ``visible'' electrons 
injected in the jet  "shell". For the primary spectrum of
$\gamma$-rays we assumed $E^{-1.5}$ with exponential cut off at $10^{19}$ eV, 
which allows to avoid the large energy release at small distances. Luminosity 
of the primary photon beam is $10^{44}$ erg/s.}
\label{fig:evolution}
\end{figure}

So far we have assumed that the 
synchrotron background in the jet
is negligible compared to the external CMBR background.  
This seems to be the case for the jet in PKS 1127-145 (see Fig. \ref{fig:PKS})
because  the radio power of the jet is small \cite{pks}.
However, as one can see from Fig. \ref{fig:spectrum}, the  densities 
of CMBR and synchrotron photons can be of the same order, as, for example in
the knot A of the jet in 3C 273 \cite{marshall}. Therefore, the 
interactions of the highest energy  
$\gamma$-rays with the radio synchrotron 
radiation  may well dominate,  at least in the 
bright knots. Besides, the random B-field in the knots can 
be stronger than in  the rest of the jet, which would result 
in increase of $E_{\rm crit}$  at which 
the electron trajectories are randomized. 
To demonstrate this effect  we performed numerical calculations
assuming very hard $\gamma$-ray spectrum with energy concentrated
around  $10^{17} \ \rm eV$.    
In Fig. \ref{fig:evolution} we  show the evolution of rate of 
production of electrons in the knot, which determines the 
synchrotron luminosity of the "shell", together with the 
profile of the synchrotron X-radiation radiation 
measured by Chandra. 
\vskip-0.8cm
\vskip-0.4cm
One can see that in the initial part of the jet, 
where synchrotron background is quite low, the 
rate of injection of the photoproduced electrons 
into the  "shell" is small. But in the knot A   
the rate  increases rapidly. An observer who detects 
the jet from a side and does not see the "main stream" of the
cascade,  may conclude  that electrons are effectively 
accelerated in the knot.

Apparently,  the formation of knots in the suggested model is a 
nonlinear process. The increase of synchrotron luminosity at some point 
of the jet would lead to an increase of the rate of ejection of electrons from the 
"main stream". This, in its turn leads to the further increase of synchrotron 
luminosity and formation of bright knots.  Note that 
although the radio luminosity of the jet is directly related to 
the injection rate of electrons in the "shell", and therefore to the 
VHE $\gamma$-ray luminosity of the central source, it depends on some
other factors as well, like the time of ``operation'' of the central 
source, the strength of the magnetic field in the jet, the escape of
electrons from the jet, etc. While the X-ray data tell us about the 
VHE $\gamma$-ray luminosity of the central source at the {\em present epoch},
the radio data rather reflect the history of evolution of the 
the central source.  
An important question is whether  photons with energies 
larger than $10^{16}$ eV  can be {\em effectively  produced} 
and at the same time {\em freely escape} the dense photon 
environments in the central engines of AGN
which are believed to be powerful particle accelerators up to   
$E_{\rm max} \ge 10^{19}$ eV \cite{acceleration}.
\vskip-0.8cm
\begin{figure}[t]
\psfig{file=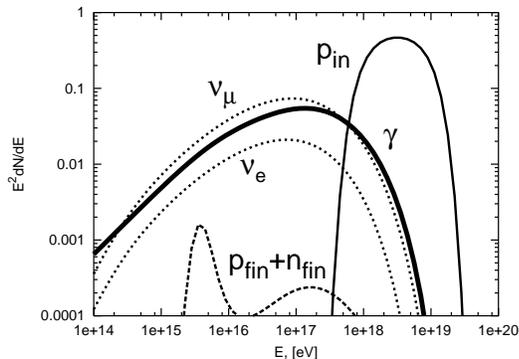,angle=-90,width =0.8\linewidth}
\caption{The spectra
of photons (thick solid line), nucleons (dashed line) and neutrinos 
(dotted lines)
after they escape the region of proton  accelerator close to 
the base of the jet. The initial proton spectrum
is shown by the thin solid line. It is assumed that 
a beam of protons is injected 
to the region of  size $10^{15} \ \rm cm$    
filled with blackbody radiation 
with temperature $10^4$ K. Total energy emitted in the VHE $\gamma$-ray beam 
is 27 \% of the energy contained in the primary photon beam normalized to 
$\int (Edn/dE)dE=1$.} 
\label{fig:core}
\end{figure}
\vskip-0.4cm
The accelerated protons can produce VHE $\gamma$-rays  
interacting with the ambient photon fields 
(supplied, for example  by accretion disk around the massive black hole) 
through photo-meson process.
Since we need a {\em beamed} $\gamma$-radiation emitted at a small 
angle to the MHD jet, the protons should cross the 
photon field almost rectilinearly. Therefore, the condition 
of a high {\em proton-to-gamma}  conversion efficiency  in the 
production region of a linear size $R$  implies    
$\tau_{p\gamma} =\sigma_{p\gamma} R ~ n_{\rm ph} \geq 1$.
On the other hand, the  produced 
$\gamma$-rays can effectively escape the production region 
if  $\tau_{\gamma\gamma} =\sigma_{\gamma\gamma} R  ~ n_{\rm ph} < 1$.
Thus, the production region can freely
escape only those $\gamma$-rays for which 
$\sigma_{\gamma\gamma}<\sigma_{p\gamma}$. 
This is possible for very energetic $\gamma$-rays 
in a ``hot''  ambient photon gas. 
As an example, we show in  Fig.~\ref{fig:core} the spectra of 
protons and $\gamma$-rays  emerging from a source  
filled with thermal radiation with $T=10^4$ K which serves a target 
for protons within a narrow energy interval between $\sim 10^{18}$ and 
$10^{19}$ eV.  Since the center of mass energy of $\gamma\gamma$ interaction
significantly exeeds the threshold of pair production, 
$(E {\rm kT})/m_{\rm e}^2 c^4 \sim 10^6 E_{17}$, 
at $E \geq 10^{17}$ eV  the pair-production  cross-section 
becomes less than  the (almost energy-independent) photo-meson 
production  cross-section, $\sigma_{p\gamma}\simeq 10^{-28}\ {\rm cm}^2$. 
Therefore the $E \geq 10^{17}$ eV 
$\gamma$-rays are not only effectively
produced, but also are able to escape the source without significant losses.
The numerical calculations shown in Fig.~\ref{fig:core} illustrate
this possibility. Note that a broader, e.g. power-law type proton
spectrum would result in  effective production of  less energetic,
$E \leq 10^{16} {\rm eV}$ $\gamma$-rays as well. 
However,  due to the increase of the $\gamma \gamma$  cross-section, 
only a small fraction of these photons can escape the source.

To conclude,  the electromagnetic cascade initiated by VHE photons 
interacting with ambient radiation  fields in the large scale 
extragalactic jets  is  an attractive  mechanism for production
of ultra-relativistic electrons (with almost 100 percent efficiency)
which can be responsible  for the observed radio-to-X-ray 
spectra of jets. The trajectories of electrons with energies below 
$E_{\rm crit}$  are isotropized by the random  magnetic field 
$B_r$. Such electrons form a "shell" around the 
cascade. Observer, who looks at the jet from a side, sees 
synchrotron and inverse Compton radiation only form the "shell"
electrons. The cascade can be developed effectively 
in the jet provided that the strength of the  
random B-field does not exceed $1 \ \mu \rm G$.
On the other hand, when  at a very large distances from
the central source the random field is reduced to a very low level
comparable with the intergalactic field,  $B \sim 10^{-9}$ eV or less,
the cascade continues almost rectilinearly until the 10-100 TeV
$\gamma$-rays start to interact effectively with the 
diffuse infrared background photons. This interactions would 
lead to the formation of {\em observable} giant pair halos with 
specific angular and energy distributions depending on the intensity
of diffuse infrared background at the  cosmological epoch corresponding 
to the redshift of the central source $z$ \cite{halo}. And finally,
if the central source is a blazar, i.e. the jet is pointed to the
observer, we may expect beamed $\gamma$-ray emission with
a characteristic for cascades $E^{-1.5}$ type spectrum extending to 
100 TeV. However, due to significant intergalactic  absorption, 
the $\gamma$-rays will arrive with significantly distorted spectra.
The possible implications of this mechanism for the  TeV blazars
like Mkn 421 and Mkn 501 are discussed elsewhere.

A.N. would like to thank SFB 375 der DFG for financial support.

\end{multicols}

\end{document}